\documentclass[prl,reprint,twocolumn, showpacs,superscriptaddress,floatfix]{revtex4-1}
\usepackage{epsfig}
\usepackage{amsmath}
\usepackage{amssymb}
\usepackage{amsfonts}
\usepackage{mathptmx}
\usepackage{dcolumn}
\usepackage{eucal}
\usepackage{bm}
\usepackage{color}
\usepackage[colorlinks,linkcolor=blue,citecolor=blue]{hyperref}
\usepackage{graphicx}
\usepackage{epstopdf}

\begin{document}

\title{Quantum Interference Hybrid Single-Electron Transistor}

\author{Emanuele Enrico}
\email{e.enrico@inrim.it}

\affiliation{INRIM, Istituto Nazionale di Ricerca Metrologica, Strada delle Cacce
91, I-10135 Torino, Italy}

\homepage{www.inrim.it}

\author{Elia Strambini}

\affiliation{NEST, Istituto Nanoscienze-CNR and Scuola Normale Superiore, Piazza
S. Silvestro 12, Pisa I-56127, Italy}

\author{Francesco Giazotto}
\email{francesco.giazotto@sns.it}

\affiliation{NEST, Istituto Nanoscienze-CNR and Scuola Normale Superiore, Piazza
S. Silvestro 12, Pisa I-56127, Italy}


\maketitle
\textbf{The manipulation of single charge quanta in solid state systems has been a field of active interest for several decades~\cite{GrabertDevoret1992,PhysRevB.53.13682}.
Thanks to Coulombic repulsion between electrons, single charges can be confined in small metallic islands yielding to the so-called Coulomb-blockade regime \cite{AverinLikharev1986}.
Many quantum phenomena, such as the tunneling effect \cite{Pekola2008} or the Josephson coupling  in superconducting junctions \cite{Giazotto2011}, have been exploited so far to control the charging state of a conducting island, and enable a number of applications ranging from on-chip cooling \cite{PhysRevLett.99.027203,PhysRevLett.103.120801,PhysRevLett.98.037201,RevModPhys.78.217} and current generation \cite{PekolaRevModPhys2013,Giblin2012, POTHIER1991573} to single-photon detection \cite{app6020035}. 
Yet, in all Coulombic systems the external control parameters play a crucial role and their exploitation requires sometimes a rather complex experimental configuration~\cite{KellerAPL,PhysRevB.64.235418,PhysRevLett.91.177003,PhysRevLett.100.177201,VartiainenAPL}.\\
Here we show that quantum interference established in a superconducting nanowire \cite{Giazotto2010} can act as a control parameter providing a phase-tunable energy barrier which allows charge manipulation with enhanced functionalities \cite{Enrico2016}. 
This additional degree of freedom for single electronics stems from the interplay between magnetic flux-dependent proximity effect \cite{GiazottoHeatModulatorAPL,PhysRevApplied.2.024005,Heikkila2002} and charging levels discretization existing in a Coulomb-blockaded
island.
Our quantum interference nanostructure represents the first realization of an innovative superconducting hybrid single-electron transistor called SQUISET \cite{Enrico2016} in which the charge landscape is phase-coherently manipulated by an external magnetic flux.
The interferometric nature of the transistor adds new perspectives to single electronics making it a promising building block in quantum metrology \cite{Flowers1324}, coherent caloritronics~\cite{MartinezPerezAPL,GiazottoHeatModulatorAPL,MartinezPerez2015quatris},
and quantum information technology \cite{Quantum2010}.}

When a short metallic nanowire (i.e., a weak-link) interrupts a superconducting (S) ring its electronic density of states (DOSs) is strongly modified by the proximity effect. The latter occurs owing to the intimate contact with the superconductor, and turns out to be controllable via a magnetic flux piercing the loop. 
This property is at the basis of the superconducting quantum interference proximity transistor (SQUIPT)~\cite{DAmbrosio2015,Giazotto2010,PhysRevApplied.2.024005,Strambini2016}. 
In this context, the behavior of a fully-superconducting SQUIPT, i.e., where the nanowire itself  is a superconductor, has been recently theoretically investigated \cite{Virtanen2016}.
Specifically, it has been shown that when the nanowire length is smaller or comparable to the superconducting coherence length its DOSs is characterized by a flux-tunable BCS-like energy gap, $\Delta\left(\Phi\right)=\Delta_0\left|\mathrm{cos}\left(\pi\Phi/\Phi_{0}\right)\right|$, where $\Delta_0$ is the zero-temperature energy gap of the superconducting ring, $\Phi$ the magnetic flux piercing the loop, and $\Phi_{0}\simeq 2\times 10^{-15}$ Wb is the flux quantum. 
The above functional behavior for the gap has been proposed as the building block of an innovative Coulombic turnstile \cite{PhysRevLett.64.2691} called superconducting quantum interference single-electron transistor (SQUISET) \cite{Enrico2016}.
In the SQUISET, the charging state of a metallic island, governed by Coulomb blockade, can be controlled by manipulating the DOSs of its source and drain electrodes made of two independent SQUIPTs. 
This defines a unique superconducting single-electron (SE) transistor in which SE injection occurs thanks to an external magnetic flux instead of the widespread electrostatic gating.

\begin{figure*}[ht]
\begin{centering}
\includegraphics[width=1\textwidth]{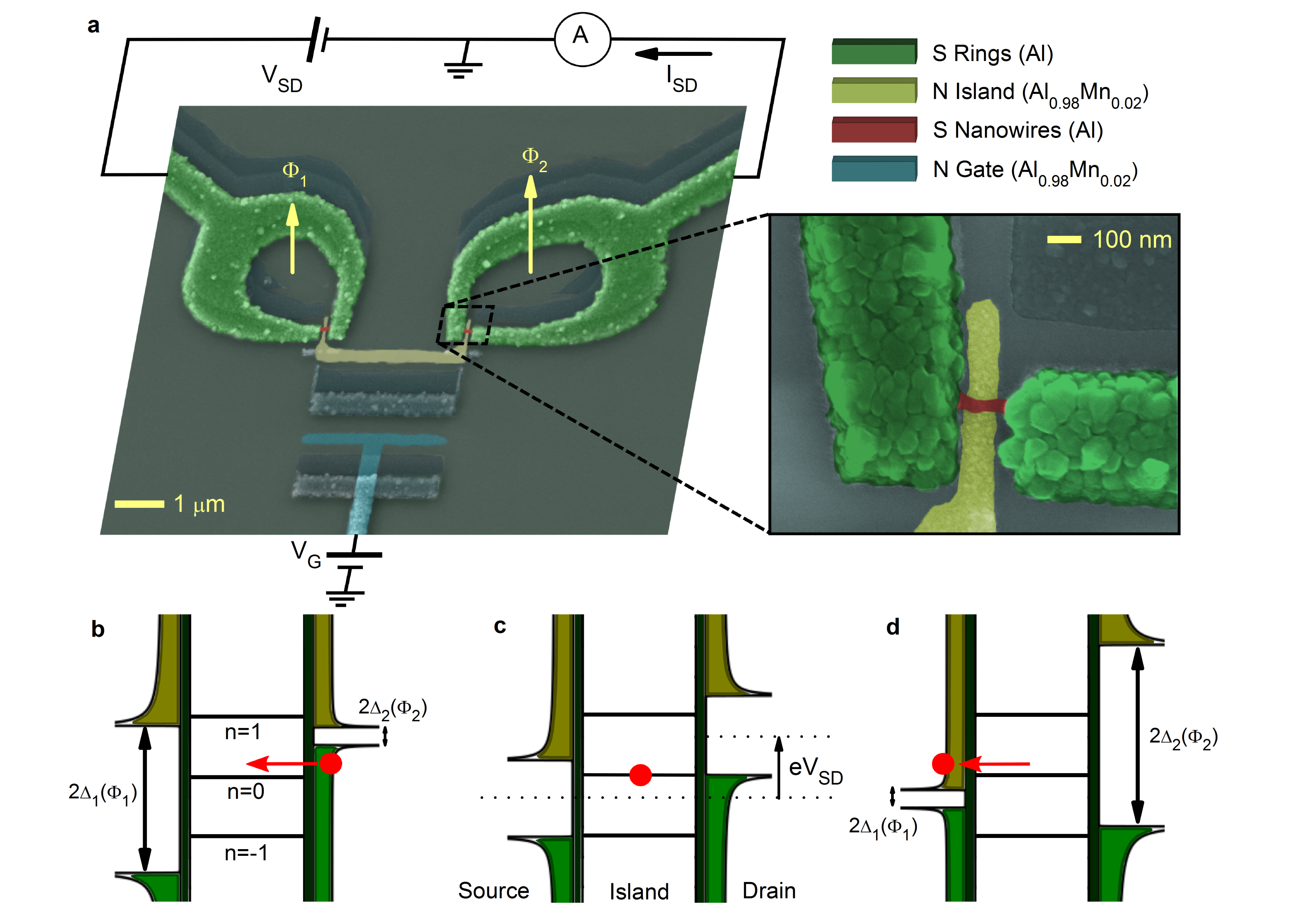}
\par\end{centering}
\caption{\textbf{The SQUISET: a single-electron transistor based on active source-drain superconducting proximity interferometers}
\textbf{a}, Pseudo-color scanning-electron micrography of the SQUISET. Two S loops made of aluminum (Al, green) are interrupted by two superconducting nanowires (Al, red), as shown in the blow-up. 
The magnetic fluxes piercing the structure ($\Phi_{1}$ and $\Phi_{2}$) are different as a result of different rings areas in a uniform magnetic field.
The island (Al$_{0.98}$Mn$_{0.02}$, yellow) is coupled to the nanowires through oxidized tunnel junctions of normal-state resistance $R_{T}\approx63$~k$\Omega$ for both  weak-links, and is capacitively-coupled to the gate electrode (light blue). 
\textbf{b}-\textbf{d}, Energy band diagrams representing a turnstile cycle composed by a single-electron (SE) injection driven by the two magnetic fluxes: SE loading from the drain electrode into the island (b), SE holding caused by Coulomb blockade (c), SE unloading from the island into the source electrode (d).
\label{fig1:Device}}
\end{figure*}

Here we report the first realization of the SQUISET and its phase-coherent magneto-electric characterization demonstrating its compatibility with a one-control parameter turnstile cycle. 
The SQUISETs are fabricated by standard three-angle shadow-mask~\cite{PhysRevLett.59.109} deposition of metals through a conventional suspended resist mask (see Methods). A pseudo-color scanning electron micrography of the core of one typical device is shown in Fig. \ref{fig1:Device}a.
Two $150$-nm-thick superconducting aluminum (Al) rings of different areas $A_{1}$ and $A_{2}$ are interrupted by two Al nanowires with length $L\approx 150$~nm, thickness $t\approx 20$~nm and width $w\approx 35$~nm. 
At low temperature and zero magnetic field the energy gap in both nanowires is $\Delta_0 \approx 198$~$\mu$eV.

\begin{figure*}[ht!]
\begin{centering}
\includegraphics[width=1\textwidth]{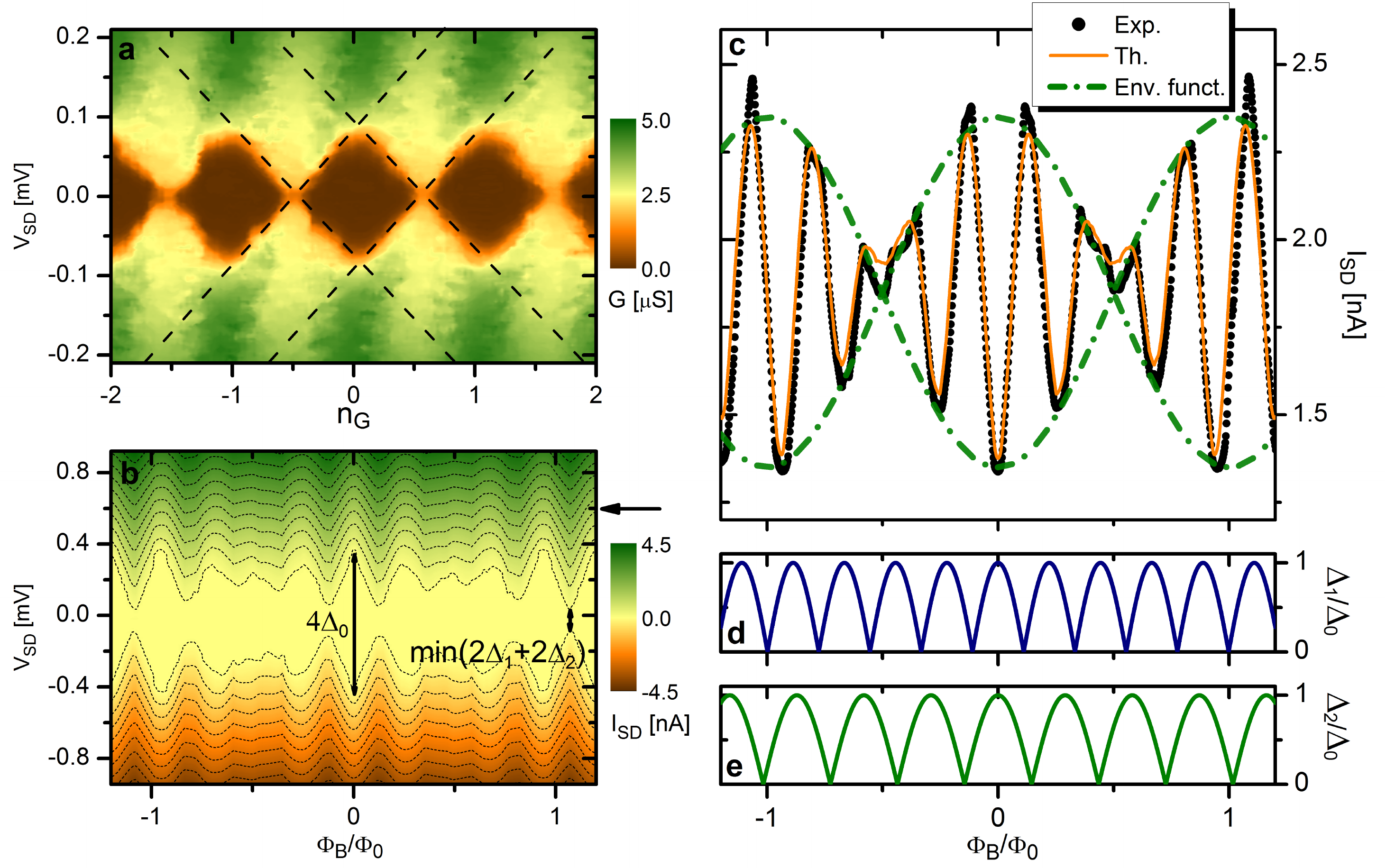}
\par\end{centering}
\caption{\textbf{Preliminary characterization of the SQUISET}
\textbf{a}, Source-drain conductance ($G$) stability diagram  recorded at a high magnetic field ($50$ mT) which quenches all the superconducting elements of the transistor. From the Coulomb-diamond diagram we estimate the transistor charging energy, $E_{C}\simeq 48$ $\mu$eV.
\textbf{b}, Source-drain  current $I_{SD}$ vs $\Phi_{B}$ and $V_{SD}$ at $n_{G}=0.5$. The blockaded region is tuned by the magnetic flux leading to a beating-like modulation. All  measurements were taken at 21 mK of bath temperature. The black arrow indicates the $V_{SD}$ condition at which the trace in panel (c) has been recorded.
\textbf{c}, Source-drain current ($I_{SD}$) characteristic vs $\Phi_{B}$ at $n_{G}=0.5$ and $V_{SD}=0.6$ mV (indicated by the horizontal black arrow in Fig.~\ref{fig2:PreliminaryMeasurements}b. The theoretical curve confirms the nearly optimal agreement with the experiment.
The envelope functions shown as dot-dashed  lines are given by $I_{SD}=\pm I_{B}\cdot \cos\left(\pi\Phi_{B}/\Phi_{0}\right)+I_{0}$,
where $I_{B}=0.5$ nA and $I_{0}=1.85$ nA. 
\textbf{d}-\textbf{e},
Theoretical behavior of the superconducting energy gaps $\Delta_1$ and $\Delta_2$ in the nanowires vs beating magnetic flux ($\Phi_{B}$).
The parameters used for the calculations are extracted from the fitting of the experimental data shown in panel (c) (see Methods).
\label{fig2:PreliminaryMeasurements}}
\end{figure*}

These nanowires realize the flux-tunable source and drain electrodes of the transistor, and are controlled by a single magnetic field ($B$) piercing the two S loops of different areas thereby yielding different magnetic fluxes $\Phi_{1}= B A_1$ and $\Phi_{2}=B A_2$. 
The SE transistor is completed by a normal metal island (N) capacitively-coupled to a gate electrode, and electrically connected to the source-drain nanowires via two  AlOx tunnel junctions with a normal-state resistance $R_{T}\approx63$ k$\Omega$. 
In the operating conditions, the SQUISET is voltage biased across the source-drain electrodes ($V_{SD}$), and the gate capacitor is polarized with $n_{G}=C_{G}V_{G}/e$ elementary charges, where $V_G$ is gate voltage, $C_{G}$ is the gate-island capacitance, and $e$ is the electron charge. 
In this configuration a discrete number of SEs can be injected from the drain into the source electrode by operating the two magnetic fluxes $\Phi_{1}$ and $\Phi_{2}$. Figures~\ref{fig1:Device}b-d sketch a typical clocking sequence illustrating the working principle of this SE transistor. 
A SE is first transferred from the drain into the island (Fig.~\ref{fig1:Device}b) where it is spatially localized owing to the finite charging energy (Fig.~\ref{fig1:Device}c) after which it is released into the source electrode (Fig.~\ref{fig1:Device}d).

The charging energy of the island, the crucial parameter for SE protection, is first characterized independently from all the other elements of the SQUISET by quenching superconductivity in the device with a magnetic field $B=50$~mT (i.e., $B> B_c$ where $B_c$ is the Al critical field). 
In this condition the SQUISET behaves as a conventional SE transistor made of normal metal source-drain electrodes connected to a Coulombic island.
From the "ortodox" analysis of the stability diagram shown in Fig. \ref{fig2:PreliminaryMeasurements}a we estimate the system charging energy $Ec= e^2/(C_G+C_S+C_D)\simeq 48$ $\mu$eV, the gate capacitance $C_G = 0.325$ fF, and the source and drain capacitances ($C_S \simeq C_D \simeq 1.5 $ fF). 
The nearly symmetric Coulomb-blockade diamonds of the diagram and the resulting source and drain capacitances confirm the similarity between source and drain tunnel junctions, an important condition for the optimal functionality of the SQUISET~\cite{Enrico2016}.
At lower field ($B < B_c$), superconductivity is recovered in the source and drain electrodes leading to an extension of the  blockade region of $2(\Delta_1+\Delta_2)$, where $\Delta_1$ and $\Delta_2$ are the energy gaps in the two nanowires. This appears in Fig.~\ref{fig2:PreliminaryMeasurements}b showing the evolution of $I_{SD}$ on $V_{SD}$ and $B$. 
Here, we have conveniently normalized $B$ by introducing the beating magnetic flux, $\Phi_B = B(A_2-A_1)/2$, which describes the behavior of the transistor in the magnetic field, i.e., beating-like oscillations of the blockade regions and $I_{SD}$, as clearly visible in the envelope functions displayed in Fig. \ref{fig2:PreliminaryMeasurements}c.
A frequency analysis of these oscillations allows to deduce the loops areas, $A_{1}\simeq5$ $\mu$m$^2$ and $A_{2}\simeq6.5$ $\mu$m$^2$ (see Methods).

With the aid of the above quantities and the theoretical model for the SQUIPT~\cite{Virtanen2016} we can determine the flux-dependent behavior of the energy gaps in the two superconducting nanowires~(see Figs. \ref{fig2:PreliminaryMeasurements}d-e).
Then, the source-drain current flowing through the SQUISET ($I_{SD}$) is obtained through the ``ortodox'' theory of quantum tunneling (see Methods)~\cite{Enrico2016}. 
The good agreement between the theoretical and the experimental $I_{SD}$ shown in Fig.~\ref{fig2:PreliminaryMeasurements}c confirms the validity of our analysis

\begin{figure*}[ht]
\begin{centering}
\includegraphics[width=1\textwidth]{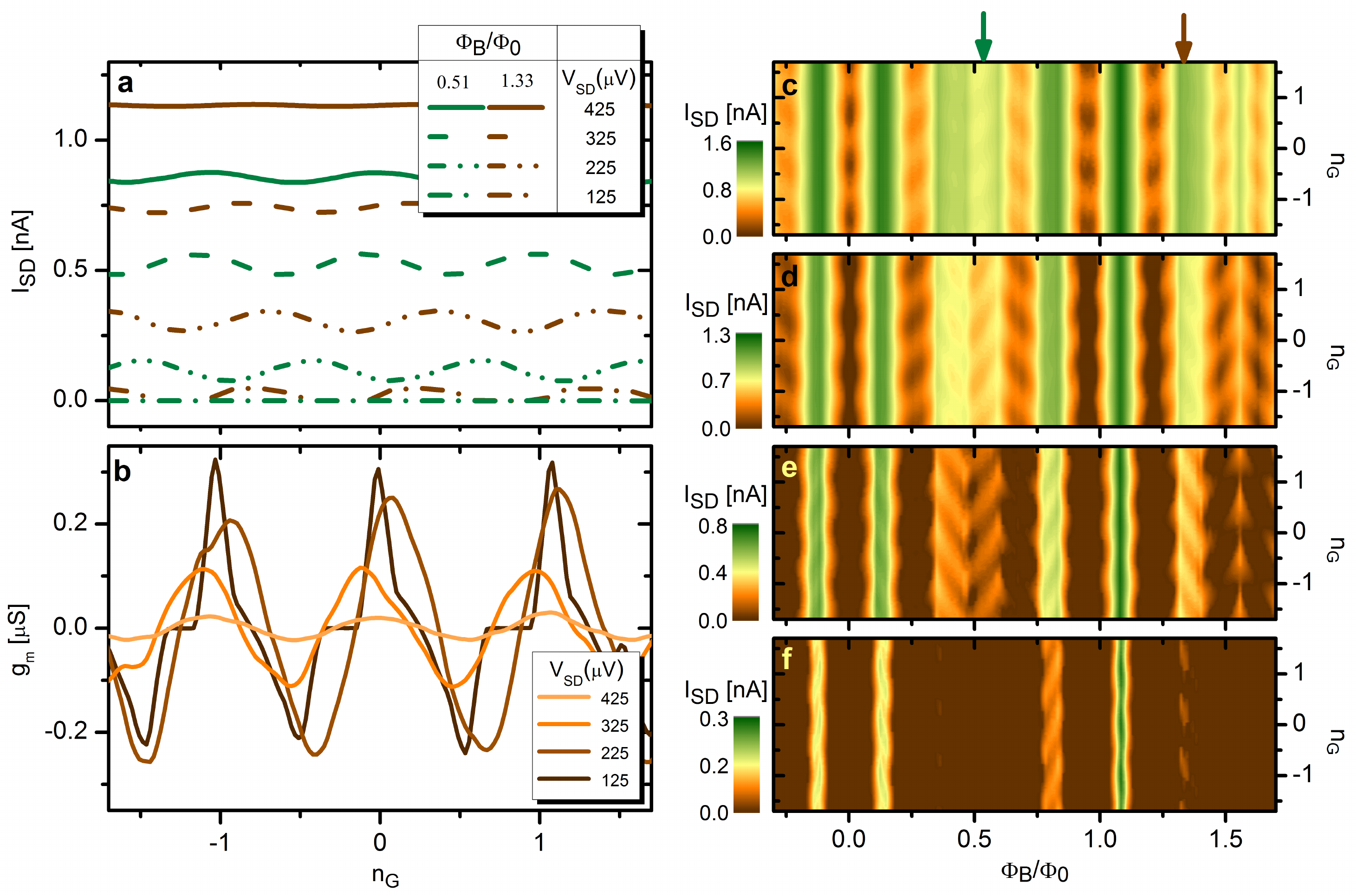}
\par\end{centering}
\caption{\textbf{Coulomb oscillations in the presence of a magnetic field.}
\textbf{a}, Selection of four representative source-drain currents $I_{SD}$ vs $n_G$ recorded at different $V_{SD}$ for two magnetic flux values. Green(brown) lines correspond to $\Phi_{B}/\Phi_{0}=0.51$($1.33$). These two flux conditions are highlighted by arrows with corresponding color on top of panel (c). 
\textbf{b}, Electrical transconductance characteristics $g_{m}=\partial I_{SD}/\partial V_{G}$ vs $n_G$  recorded for selected values of $V_{SD}$ at $\Phi_{B}/\Phi_{0}=1.33$.
\textbf{c}-\textbf{f}, Contour plot of source-drain current $I_{SD}$ vs $\Phi_{B}$ and $n_G$ for selected values of $V_{SD}$: 425~$\mu$V (c), 325~$\mu$V (d), 225~$\mu$V(e), and 125$\mu$V (f). These plots show how the magnetic flux can influence the energy scale in which the transistor is sensitive to gate voltage variations. 
All  measurements were taken at 21 mK of bath temperature.
\label{fig2:Oscillations}}
\end{figure*}

Having identified independently the role of $n_{G}$ and $\Phi_{B}$ in the Coulomb-blockade operation of the transistor (see Figs.~\ref{fig2:PreliminaryMeasurements}a,b), we now investigate the interplay between these two physical quantities for the control of the island charging state.
In conventional Coulombic islands the latter is usually characterized through Coulomb oscillations (see Fig.~\ref{fig2:Oscillations}a and b), where an imposed source-drain voltage ($V_{SD}$)  define the energy band for the island charge states which, controlled by the gate voltage, transports the current through the transistor.
As shown before, in our device this energy band is reduced by the flux-tunable superconducting energy gaps as a result of their beating-like behavior in $\Phi_{B}$ which modulates as well the Coulomb oscillations.
The role of the gate control parameter is confirmed by Fig.~\ref{fig2:Oscillations}b where, for a chosen condition ($\Phi_{B}=1.33\Phi_{0}$) for which $\Delta_{1}$ and $\Delta_{2}$ are small and almost in phase, the transconductance is maximized around integer values of $n_{G}$, and minimized for $n_{G}$ half-integer values. 
As a consequence of the small value of $\Delta_1+\Delta_2$, the smaller is $V_{SD}$ the more pronounced is the gate effect.

This interplay is clearly shown in Figs.~\ref{fig2:Oscillations}c-f which display the Coulomb oscillations vs $n_{G}$ and $\Phi_{B}$ for selected values of source-drain voltage ($V_{SD}=425$~$\mu$V, $V_{SD}=325$~$\mu$V, $V_{SD}=225$~$\mu$V, and $V_{SD}=125$~$\mu$V from top to bottom, respectively).
For large values of $V_{SD}$ (see Fig. \ref{fig2:Oscillations}c), source-drain current is not blocked in the whole space of parameters apart from small lobes (two-dimensional analogue of the Coulomb peaks) located at $\Phi_{B}=0$ with $\Phi_{0}$ periodicity, and unitary periodicity in $n_{G}$.  In these conditions the two superconducting energy gap are in-phase at their maximum while the gate voltage modulates the current from a blockaded (brown) to a conducting region (orange).
At lower bias voltages (see Fig. \ref{fig2:Oscillations}d-f) extended plateaus of zero current appear nearby $\Phi_{B}\sim 0$ and $\Phi_{B}\sim\Phi_{0}$ due to the reduced available energy which is not sufficient to transfer any electron across the tunneling barriers. Accordingly, the Coulomb lobes are shifted in magnetic flux reaching the regions of $\Phi_{B}=0.5 \Phi_{0}$ and $\Phi_{B}=1.5 \Phi_{0}$ for $V_{SD}=225$~$\mu$eV. These are the relevant regions that experimentally demonstrate the capability of our device to operate in a magnetic-flux subspace condition, where the two energy gaps $\Delta_1(\Phi_1)$ and $\Delta_2(\Phi_2)$ are almost in anti-phase (see Fig. \ref{fig2:PreliminaryMeasurements}d-e) with respect to their own magnetic fluxes ($\Phi_2-\Phi_1=\Phi_{0}/4 $).
At very low bias ( Fig. \ref{fig2:Oscillations}f) the current is still finite for some specific magnetic-field values, with a maximum occurring at $|\Phi_{B}|=1.08\Phi_{0}$ which corresponds to the value that minimizes the sum $\Delta_1+\Delta_2$ (see Fig. \ref{fig2:PreliminaryMeasurements}d,e). 
This value is consistent as well with the minimum observed in the stability diagram of Fig. \ref{fig2:PreliminaryMeasurements}b.

\begin{figure*}[ht]
\begin{centering}
\includegraphics[width=1\textwidth]{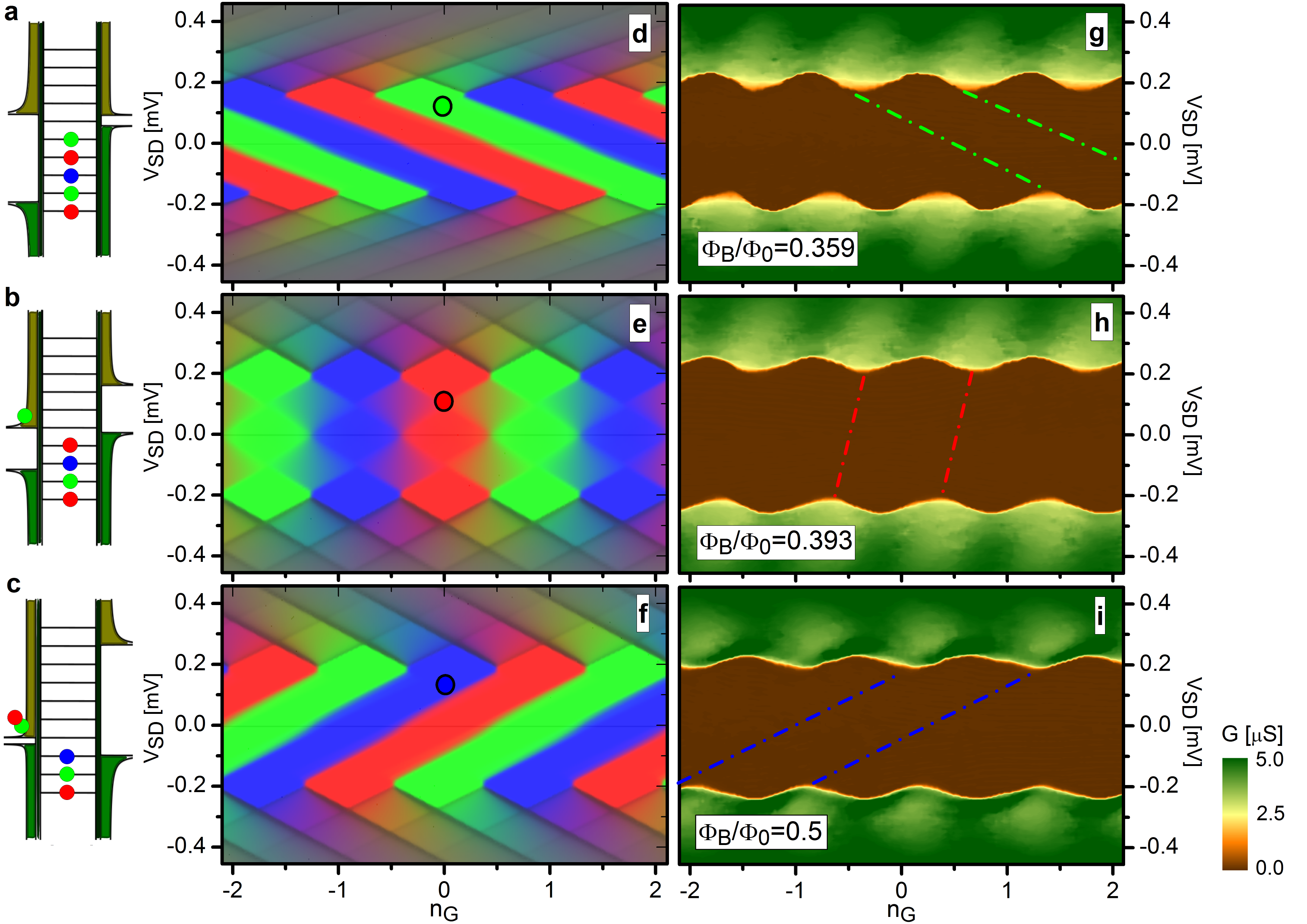}
\par\end{centering}
\caption{\textbf{Multi-electron turnstile behavior} 
\textbf{a}-\textbf{c},Energy band diagrams of three representative electro-magnetic configurations of the transistor. Here we set $n_{g}=0$ and $eV_{SD}=\Delta_{0}/2$. It clearly appears the influence of the magnetic flux on the number of excess charges on the island being in the blockade regime and at fixed bias voltage. 
The empty and the full branches of the quasiparticles DOSs are vertically shifted by the external magnetic flux loading or unloading different charging levels.
\textbf{d}-\textbf{f}, Stability diagrams showing the calculated differential conductance ($G=\frac{\partial I_{SD}}{\partial V_{SD}}$) for the three different charge configurations (see Supplementary Information Fig. S1 for more stability diagrams). RGB channels are proportional to the probability to find an electron in different island energy levels. Pure red, green and blue colors correspond to blockaded regions (Coulomb diamonds). 
Black circles represent the configurations depicted in the diagrams of panels (a-c) and correspond to the selected charge states indicated by the dot-dashed lines in panels (g-i).
\textbf{g}-\textbf{i}, Corresponding measured differential conductance stability diagrams. 
Plots (a), (d) and (g) were obtained for $\Phi_{B}=0.359 \Phi_{0}$. In (b), (e) and (h) we set $\Phi_{B}=0.393 \Phi_{0}$ whereas in (c), (f) and (i) we set 
$\Phi_{B}=0.5 \Phi_{0}$. All  measurements were taken at 21 mK of bath temperature.
\label{fig3:Stability}}
\end{figure*}

We now show the experimental evidence of the turnstile configuration capability of our SE transistor in the anti-phase regime (i.e., at $\Phi_{B} \simeq \Phi_{0}/2$) which represent the optimal operating condition, as shown in the sketch of Fig.~\ref{fig1:Device}b-d. 
In this specific device the charging energy is smaller than the nanowires gaps ($E_C<\Delta_0$) so that more than one single charge can be loaded and unloaded in the island per operating quasi-static cycle. 
This mechanism appears clearly from Figs.~\ref{fig3:Stability}a-c where the energy band diagrams of the island are displayed for the magnetic configurations explored.
Red-green-blue (RGB) colors are associated to each single energy level of the Coulombic island, and these are recalled in Figs.~\ref{fig3:Stability}d-f representing the calculated stability diagrams for the probability to have one electron in the last occupied energy level. 
Pure RGB colors denote regions where island charge states are blocked and thereby suitable for a high-fidelity SE turnstile operation. 
Color mixing identifies those situations where the charge configuration of the island is not fixed, and a `leakage' current can flow through the transistor.

The experimental confirmation of this physical picture is provided by the stability diagrams of the differential conductance measured in these three magneto-static configurations (see Figs.~\ref{fig3:Stability}g-i). 
The diagrams show a specific skewing of Coulomb diamonds which evolves continuously as a function of the magnetic flux, in full agreement with the calculations
 (see Figs.~\ref{fig3:Stability}d-f), and stemming from the asymmetry of the energy gaps in the nanowires ($\Delta_{1} \neq \Delta_{2}$). 
Within this flux sequence the transistor can therefore operate as a SE turnstile by fixing the bias voltage in the blockade region of interest (e.g., at $V_{SD}=\Delta_0/2e$, as indicated by the black circles in Figs.~\ref{fig3:Stability}d-f), and by sweeping the external magnetic flux. 
In the present quasi-static situation, the sequence starts at $\Phi_{B}=0.359\Phi_{0}$ by loading five electrons into the island (Fig.~\ref{fig3:Stability}a), and ends at $\Phi_{B}=\Phi_{0}/2$ where two electrons are unloaded from the island. This specific example demonstrates the possibility to control the transfer of two electrons from drain to source without using the gate voltage. 
By experimental evidence it is clear how a magnetic flux cycle allow the control of a discrete number of electron flow between the source and the drain electrode.

We wish to further stress the role embodied by the magnetic flux as a control parameter: our experimental results prove indeed that a SQUISET having a rather simple design can reach the turnstile capability for single charge quanta with a suitable magnetic-like gating.  

In summary, we have realized the first superconducting quantum interference single-electron transistor (SQUISET) where the magnetic flux degree of freedom provides an additional control over single-electron charge transfer. 
The transistor can be realized with conventional nanofabrication techniques, and can be integrated with on-chip superconducting coils for independent magnetic flux control, fast turnstile operation and enhanced functionalities.
Operating the device with a high-frequency electromagnetic field will provide a promising route towards the metrological definition of the current standard.
In this view the transport physics of the device proposed is, at first order, comparable to the well-known SINIS \cite{Pekola2008} turnstile with which it shares the possible pumping error sources. 
Moreover, the detrimental effects of the background charges fluctuations \cite{PhysRevB.53.13682} on the accuracy of metallic turnstiles could potentially be reduced in our SQUISET respect to other electrically gated coulombic structures since the 
two-level systems responsible for such fluctuations should be quiet under magnetic time-dependent excitation.
The SQUISET might have impact in cryogenic micro- and nanoelectronics as well as in solid-state quantum information technologies.
Extending the concept here presented to a fully superconducting structure, the charge configuration of a superconducting island could be magnetically adiabatically tuned exploiting its Josephson coupling \cite{PhysRevLett.91.177003, Quantum2010} with two SQUIPT-engineered electrodes \cite{Virtanen2016}.
Yet, combined with superconducting circuits, it could act as a building block of future coherent single nanoelectronics.

\section*{Methods}

{\textbf{Experimental} - One step of electron-beam lithography was used to
fabricate the transistors followed by a three-angle shadow-mask evaporation
of  metallic thin films through a suspended resist mask. 
The latter, lying
onto an oxidized Si substrate, was processed in a UHV electron-beam
evaporator. 
A $15$-nm-thick layer of Al$_{0.98}\text{Mn}_{0.02}$ was initially
deposited, the chip being tilted at an angle of $37{^\circ}$, to realize
the normal metal island structure. 
The sample was then exposed to
$4\times10^{-2}$mbar O$_{2}$ for $5$ min to form the AlO$_{\text{x}}$
thin insulating layer composing the tunnel junctions. 
Then the chip
was tilted to $20{^\circ}$, and a $25$-nm-thick layer of Al was
deposited to form the superconducting nanowires. 
Finally,
a $150$-nm-thick layer of Al was deposited at $0{^\circ}$ to
realize the superconducting rings.  The source and the drain electrodes
are nominally identical differing only for the rings areas.

The magneto-electric characterization of the SQUISETs was performed down to
$21$mK in a shielded and filtered dilution refrigerator.
Current and conductance  measurements were performed with
room temperature electronics whereas the magnetic field was imposed
through a superconducting coil surrounding the sample chamber.

\textbf{Theory} 
The theoretical modeling of the SQUISET was performed in the framework of the ''orthodox theory' of single-electron tunneling
under the static regime condition. 
The sequential tunneling approach in the first order approximation leads to the charge current as a solution of the standard master equation \cite{GrabertDevoret1992}:
\begin{equation}
I_{SD}\left(\Phi_{1},\Phi_{2},E_{i,\pm}\right)=-e\sum_{n}p_{n}\left(\Gamma_{1,+}-\Gamma_{1,-}\right),
\label{eq:1(ISD)}
\end{equation}
where $E_{i,\pm}=\pm E_{c}\left(n-n_{G}\pm1/2\right)\pm eV_{SD}/2$ are the free energy variations associated to a single 
electron tunneling event in the $i$-th junction that increase ($+$) or decrease ($-$) the number of excess charges on the island 
($n$), and $\Gamma_{1,\pm}\left(\Phi_{1},E_{1,\pm}\right)$ are the tunneling rates across the first junction.
Moreover, $p_{n}\left(\Phi_{1},\Phi_{2},E_{i,\pm}\right)$ is the probability to find the transistor in the $n$ charging state depending on its electro-magnetic environment. 
The full expressions for the above quantities can be found in \cite{Enrico2016}. 
We assumed each component of the transistor to be at thermal equilibrium
with the lattice phonons, and to reside at the base temperature of the cryostat.
Subgap quasiparticles populations in the source and drain electrodes have been modeled by assuming
a superconducting density of states smeared by a $\varGamma_{0}=10^{-5}\Delta_{0}$ Dynes parameter \cite{Dynes1984, PhysRevLett.105.026803}.
The magnetic fluxes, $\Phi_{i}=A_{i}B$, have been deduced assuming a uniform magnetic field $B$ perpendicular to the
SQUIPTs loop plane, and rings areas ($A_{i}$) coming from the geometrical estimate. 
Moreover, the two areas  were also deduced by peak analysis on the FFT trace of the current vs magnetic field signal, the 
latter revealing two primary components related to the different superconducting loops.
 
}{ \par}

\section*{References}
{
}{\par}

\section*{Acknowledgements}
{We thank S. D'Ambrosio, I. Mendes, G. Amato and L. Callegaro for useful discussions. E.E. acknoledges partial financial support from the European Metrology Research Programme (''EXL03 MICROPHOTON''). The EMRP is jointly funded by the EMRP participating countries within EURAMET and the European Union.
The work of E.S. is funded by the Marie Curie Individual Fellowship MSCAIFEF-ST No. 660532-SuperMag.
F.G. acknowledges
the European Research Council under the European
Unions Seventh Framework Program (FP7/2007-
2013)/ERC Grant agreement No. 615187-COMANCHE, and MIUR-FIRB2013-Project Coca (Grant No.
RBFR1379UX) 
for partial financial support.
}{\par}

\section*{Author contributions}
{E.E. fabricated the samples and carried out simulations. E.E. and E.S. performed the measurements and analysed the data. E.E. and F.G. conceived the experiment. E.E., E.S. and F.G. discussed the results and implications equally at all stages, and wrote the manuscript.
}{\par}

\end{document}